\documentclass[twoside,a4paper]{article}

\usepackage{amsmath,amssymb,bbding,ifsym,pifont}
\usepackage{booktabs,hhline,enumerate}
\usepackage{graphicx,color,subfigure}

\usepackage{bm,subfloat,url}

\newtheorem{algorithm}{Algorithm}

\newcommand{\tr}{\operatorname{tr}}

\begin{document}

\begin{center}
	{\Large Blind Source Separation with Compressively Sensed Linear Mixtures 
	\footnote{M. Kleinsteuber and H. Shen are with the Department
	of Electrical Engineering and Information Technology, Technische 
	Universit\"at M\"unchen, M\"unchen, Germany. 
	e-mail: (see http://www.gol.ei.tum.de). \\
	\indent Authors are listed in alphabetical order due to equal contribution. \\
	\indent Parts of this work have been presented at the Workshop \emph{Signal Processing with 
	Adaptive Sparse Structured Representations (SPARS'11)}, Edinburgh, June 2011 \cite{klei:spars11}. \\
	\indent This work has partially been supported by the Cluster of Excellence
	\emph{CoTeSys} - Cognition for Technical Systems, funded by the Deutsche Forschungsgemeinschaft (DFG).
	} } \\[4mm]
	{\large Martin~Kleinsteuber and Hao Shen } \\[4mm]
	\today \\[9mm]
\end{center}

\begin{abstract}
	This work studies the problem of simultaneously separating and 
	reconstructing signals from compressively sensed linear mixtures. 
	We assume that all source signals share a common sparse 
	representation basis.
	The approach combines classical Compressive Sensing (CS) theory 
	with a linear mixing model. 
	It allows the mixtures to be sampled independently of 
	each other. If samples are acquired in the time domain, this means 
	that the sensors need not be synchronized.
	Since Blind Source Separation (BSS) from a linear mixture is only 
	possible up to permutation and scaling, factoring out these 
	ambiguities leads
	to a minimization problem on the so-called oblique manifold.
	We develop a geometric conjugate subgradient method that scales to 
	large systems for solving the problem.
	Numerical results demonstrate the promising performance of the
	proposed algorithm compared to several state of the art methods.
\end{abstract}

\begin{center}
	\textbf{\small{Index Terms}} \\[2mm]
	Compressed sensing, blind source separation, oblique manifold, conjugate subgradient method.
\end{center}

\section{Introduction}
%
%
%
%
Recovering signals from only the mixed 
observations without knowing the priori information of both the 
source signals and the mixing process is often referred to as 
\emph{Blind Source Separation} (BSS), cf. \cite{hayk:uaf1}. 
Different BSS methods are used in various challenging data analysis applications, such as 
functional Magnetic Resonance Imaging (fMRI) analysis and 
microarray analysis.
%
%
In order to achieve reasonable performance, prominent methods, e.g.
Independent Component Analy\-sis (ICA), usually 
require a large number of observations
\cite{berm:ieeespl05}.
Unfortunately, the availability of a large amount of data samples can not be guaranteed  
in many real applications, due to either cost or time issues.

The theory of compressed sensing 
(CS), cf. \cite{dono:ieeetit06},
shows that, when a signal is \emph{sparse} 
(or \emph{compressible}) with respect to some basis, 
only a small number of samples suffice for 
exact (or \emph{approximate}) recovery. 
%
It is interesting to know that the concept of sparsity has also been 
used as a separation criterion in the context of BSS.
In \cite{zibu:nc01}, for example, a family of efficient algorithms in the probabilistic framework
are proposed. In \cite{Li:2004}, the authors investigate sparse representation of signals in an overcomplete basis in conjunction with BSS and motivate the usage of $\ell_1$-norm for a sparsity measure.

The approach presented in this work can be regarded as a generalization of
Morphological Component Analysis (MCA), cf. \cite{Elad:05} and \cite{Bobin:07}. Roughly speaking, while MCA takes advantage of the sparse representations of the signals only for separation tasks, we additionally employ the sparsity assumption for reconstruction. The resulting cost function is thus very much related 
to the ones that typically arise in MCA.
Nevertheless, the current scenario with compressively 
sensed samples has not been studied and thus differs from the existing
MCA models. Although it has the potential of tackling 
the underdetermined BSS case, our numerical validation only considers
the determined case, where the number of sources equals the number of mixtures.

By considering the classical result that the source signals can only be
extracted with scaling ambiguity, we regularize 
our problem by restricting each column of the mixing matrix to have unit norm.
In other words, the resulting optimization problem is 
defined on the so-called oblique manifold.
Furthermore, as many potential applications of the present work lie in the 
area of image or audio signal processing, any promising algorithms have to 
be capable of scaling to large systems.
It is well known that conjugate gradient-type methods provide a nice 
trade-off between large scale problems and good convergence behavior.
In this work, we propose an approach that lifts the
conjugate subgradient method proposed by Wolfe in \cite{cg:wolfe:1975} 
to the manifold case.

The paper is outlined as follows. 
Section~\ref{sec:02} introduces some basic concepts and provides a 
description of the compressively sensed BSS problem. 
In Section~\ref{sec:03}, we develop a geometric subgradient algorithm.
Finally, numerical results and conclusions are
given in Section~\ref{sec:04} and Section~\ref{sec:05} respectively.

\section{Problem Description}                      \label{sec:02}
\subsection{Notation and Setting}
For the sake of convenience of presentation, we follow the notation in the compressive sensing literature and
represent the signals as column vectors.
The instantaneous linear BSS model is given by
%
\begin{equation}
\label{eq:01}
	Y = S A,
\end{equation}
where $S = [s_{1}, \ldots, s_{m}] \in \mathbb{R}^{n \times m}$ 
denotes the data matrix of $m$ sources with $n$ samples ($m \ll n$), 
$A = [ a_{1}, \ldots, a_{k} ] \in \mathbb{R}^{m \times k}$ 
is the mixing matrix of full rank, and $Y = [y_{1}, \ldots, y_{k}] 
\in \mathbb{R}^{n \times k}$ represents the $k$ linear mixtures of $S$.
%
%
The task of standard BSS is to estimate the sources $S$, given only 
the mixtures $Y$.

We assume that for all sources $s_{i} \in \mathbb{R}^{n}$, there exists 
an (overcomplete) basis ${\mathcal D} \in \mathbb{R}^{n \times d}$, $n \leq d$,
of $\mathbb{R}^{n}$, referred to as \emph{representation basis}, such that
$s_{i} = {\mathcal D} x_{i},$ with sparse $x_{i} \in \mathbb{R}^{d}$.
%
%
More compactly, this reads as
\begin{equation}
\label{eq:03}
	S = {\mathcal D} X,
\end{equation}
where $X = [x_{1}, \ldots, x_{m}] \in \mathbb{R}^{d \times m}$.

%

\subsection{Compressively sensed BSS model}

Now let us take one step further to compressively sample each 
mixture $y_{i} \in \mathbb{R}^{n}$ individually.
To this end, we denote the
 \emph{sampling matrix} for the mixture $y_i$ by
$\Phi_{i} \in \mathbb{R}^{p_{i} \times n}$ for $i = 1,\ldots, k$.  
Then, a compressively sensed observation $\widehat{y}_{i} \in \mathbb{R}^{p_{i}}$ of the $i$-th mixture is given by
\begin{equation}
\label{eq:04}
	\widehat{y}_{i} = \Phi_{i} y_{i} = \Phi_{i} {\mathcal D} X a_{i}.
\end{equation}
We refer to \eqref{eq:04} as the \emph{Compressively Sensed BSS} (CS-BSS) model.
To summarize, the task considered in this work is formulated as:
Given the compressively 
sensed observations $\widehat{y}_{i} \in \mathbb{R}^{p_{i}}$, for $i = 1, \ldots, k$,
together with their corresponding sampling matrices $\Phi_{i} \in \mathbb{R}^{p_{i} \times n}$, 
estimate the mixing matrix $A \in \mathbb{R}^{m \times k}$ and the sparse 
representations $X \in \mathbb{R}^{n \times m}$.

There are various measures of sparsity available in the literature, cf. \cite{hurley:09}.
In this work, we confine ourselves to the $\ell_1$-norm as a measure of sparsity, because
(i) it is suitable for numerical algorithms, in particular for large scale problems;
(ii) it is an appropriate prior for many real signals, cf. \cite{zibu:nc01}.
%
This leads to the following optimization problem
\begin{equation}
\label{eq:optprob1}
%
	\operatorname*{argmin}_{X,A} \|X\|_1,
		\;\;\, {s.t.} \;\;\, \widehat{y}_{i} = \Phi_{i} {\mathcal D} X a_{i},\; i=1,\dots,k. 
\end{equation}
In real applications, it is unavoidable that the observations $\widehat{y}_{i}$ 
are contaminated by noise. 
Hence let $\epsilon_i$ denote the error radius for the observation $i$. The optimization problem 
\eqref{eq:optprob1} turns into
\begin{equation}
\label{eq:optprob2}
%
	\operatorname*{arg min}_{X,A} \|X\|_1,
	\;\;\, {s.t.} \;\;\, \|\widehat{y}_{i} - \Phi_{i} {\mathcal D} X a_{i}\|_2 \!\leq\! \epsilon_i,\; i\!=\!1,\dots,k.
\end{equation}
Following a standard approach, we formulate \eqref{eq:optprob2} in an unconstrained Lagrangian form, namely
\begin{equation}
\label{eq:optprob3}
	\operatorname*{arg min}_{X,A} \; \|X\|_1 + \sum\limits_{i=1}^{k} \lambda_{i}
 \|\widehat{y}_{i} - \Phi_{i} {\mathcal D} X a_{i}\|_{2}^{2},
\end{equation}
where the scalars $\lambda_{i} \in \mathbb{R}^{+}$ weigh the reconstruction error
of each mixture individually according to $\lambda_i \sim 1/\epsilon_i^2$, and balance these errors against the sparsity term $\| X \|_{1}$.
It is well known that, in compressive sensing, inappropriate regularization
parameters might lead to not only slow convergence but also
local optima. To cope with this issue, we follow the adaptive 
updating strategy proposed in \cite{bobi:tip07,wrig:ieeetsp09}.
A rigorous analysis of the updating strategy in our CS-BSS setting
is beyond the scope of this work.

\subsection{Regularized CS-BSS problem}

Obviously, problem \eqref{eq:optprob3} is ill-posed. 
Indeed, an optimization procedure would let the norm of $A$ explode to drive $\|X\|_1$ to zero.
In order to regularize the problem, we therefore restrict to the case where $\|a_i\|_2 = 1$ for all $i=1, \dots, k$.
%
%
Thus, together with the full rank condition, 
we restrict the mixing matrix $A$ onto the oblique 
manifold \cite{shen:lva10}
\begin{equation}
\label{eq:06}
	\mathcal{OB}(m,k) \!:=\! \left\{ \!A \!\in\! \mathbb{R}^{m \times k} \big| \!\operatorname{rk}(A) \!=\! k, \operatorname{ddiag} (A^{\top}\! A) \!=\! I_{k} \right\}\!,
\end{equation}
where $I_{k}$ is the $(k \times k)$ identity matrix, and
$\operatorname{ddiag}(Z)$ forms a diagonal matrix, whose diagonal entries are those of $Z$.
Note, that this is a common approach in many BSS scenarios, since it is well known
\cite{como:sp94}
that the mixing matrix $A$ is 
identifiable only up to a column-wise scaling and permutation.
The regularized problem that we will consider is hence given by
\begin{equation}
\label{eq:optprob4}
	\operatorname*{argmin}_{ \substack{ X\in \mathbb{R}^{d \times m} \hspace*{4.8mm} \\ A \in \mathcal{OB}(m,k) } } 
	\|X\|_1 + \sum\limits_{i=1}^{k} \lambda_{i}
	\left\|\widehat{y}_{i} - \Phi_{i} {\mathcal D} X a_{i}\right\|_{2}^{2}.
\end{equation}
The optimization problem is thus defined on the product manifold $\mathbb{R}^{d \times m} \times \mathcal{OB}(m,k)$.

\section{A Conjugate Subgradient type method for the CS-BSS problem}
\label{sec:03}

\subsection{Conjugate Subgradient on Riemannian manifolds}
In order to tackle the non-smooth problem \eqref{eq:optprob4}, we propose a conjugate subgradient
type approach on manifolds by generalizing the method proposed in \cite{cg:wolfe:1975}.
A generalization of subgradient methods to the Riemannian manifold case has been studied in \cite{ferreira:98}. We refer to \cite{ledyaev:07} for an introduction to nonsmooth analysis on manifolds,
where the required concepts are introduced.
Generally, for minimizing a nonsmooth function $f \colon M \to \mathbb{R}$
where $M$ is a Riemannian manifold and a subgradient of $f$ exists for all $x \in M$,
we propose the following scheme, further referred to as \emph{Riemannian Conjugate Subgradient (CSG) method}. Let $T_xM$ be the tangent space at $x$, let $\partial f(x) \subset T_xM$
denote the Riemannian subdifferential of $f$ at $x$, i.e. the set of all Riemannian subgradients at $x$, 
and let $\| \cdot \|_R$ denote the norm on $T_x M$ which is 
induced by the Riemannian metric $\langle \cdot, \cdot \rangle_R$. 
Moreover, let
\begin{equation}
G(x):=\operatorname*{argmin}_{\xi \in \partial f(x)} \| \xi \|_R
\end{equation}
denote the subgradient with smallest Riemannian norm.
Let $x_0 \in M$ be an initial point for the algorithm. Denote $G_i := G(x_i)$, and
the descent direction at the $i$-th iteration by $H_i \in T_{x_i}M$. By initializing
$H_0:=-G_0$, the algorithm now iteratively updates 
\begin{equation}\label{eq:stepsize}
x_{i+1}= \exp_{x_i}\!\left( \alpha_i H_i \right),
\end{equation}
where $\exp$ denotes the Riemannian exponential and the scalar $\alpha_i$ is 
the line search parameter at the $i$-th iteration.
Various line-search techniques for computing $\alpha_i$ exist
from which we choose \emph{backtracking line-search} \cite{nocedal:2006}, cf. also \cite{absi:oamm08}
for the Riemmanian case, as it is conceptually simple and computationally cheap. 

There are also several formulas

available that update the descent direction, e.g. Hestenes-Stiefel, Polak-Ribi\`ere, and Fletcher-Reeves, which can all be generalized straightforwardly to the manifold setting. Here, we restrict to a manifold adaption of Hestenes-Stiefel. Let $\xi \in T_{x_i}M$ and 
denote by $\tau_i(\xi) \in T_{x_{i+1}}M$ the parallel transport of $\xi$ along the geodesic $\gamma(t)= \exp_{x_i}\left( t H_i \right)$ into the tangent space $T_{x_{i+1}}M$. The direction update is
then given by
\begin{equation}\label{eq:dirupdate}
H_{i+1}:= - G_{i+1} + \frac{\langle G_{i+1}, G_{i+1}- \tau_i(G_i) \rangle_R}{\langle \tau_i(H_i), G_{i+1}- \tau_i(G_i)\rangle_R} \tau_i(H_i).
\end{equation}

Equations \eqref{eq:stepsize} and \eqref{eq:dirupdate} are iterated until a stopping criterion is met. Usually, conjugate gradient methods use
a reset after $N={\rm dim} M$ iterations, i.e. $H_i:=-G_i$ if $i \mod N=0$.
We propose to reset the direction when no significant update on the search directions is observed.

\subsection{The Riemannian subgradient for CS-BSS}
%
The manifold $M=\mathbb{R}^{d \times m} \times \mathcal{OB}(m,k)$ is endowed with 
the Riemannian metric inherited from the surrounding Euclidean space, that is
\begin{equation}\label{eq:riemannianmetric}
\langle (X_1, A_1), (X_2, A_2) \rangle_R := \tr (X_1 X_2^\top)+\tr(A_1 A_2^\top).
\end{equation}
Let $\lambda_i > 0$ for $i=1, \dots, k$ and let 
%
\begin{equation}
	\begin{split}
		f \colon \mathbb{R}^{d \times m} \times & \mathcal{OB}(m,k) 
		\to \mathbb{R}, \\
		f(X,A) & = \|X\|_1 + \sum\limits_{i=1}^{k} \lambda_{i}
		\|\widehat{y}_{i} - \Phi_{i} {\mathcal D} X a_{i}\|_{2}^{2}.
	\end{split}
\end{equation}
The Riemannian subdifferential
 of $f$ at $(X,A)$ with respect to the Riemmanian metric 
 \eqref{eq:riemannianmetric} is given by
\begin{equation}
\partial f (X,A)=\{ \left(\partial_1 f (X,A), \partial_2 f (X,A) \right) \},
\end{equation}
where
\begin{equation}
\label{eq:17}
	\partial_1 f (X,A) =  \partial \|X\|_1 + \sum\limits_{i=1}^{k} \lambda_{i}  ( \Phi_{i} {\mathcal D})^{\!\top} \!\!\left(\Phi_{i} {\mathcal D} X a_{i} \!-\! \widehat{y}_i \right) a_i^\top 
\end{equation}
and
\begin{equation}
\partial_2 f (X,\!A) \!=\!\! \left[\lambda_i \Pi(a_{i}) (\Phi_{i} {\mathcal D} X)^{\!\top}\!\!\left(\Phi_{i} {\mathcal D} X a_{i} \!-\! \widehat{y}_i \right) \right]_{i=1,\dots,k}\!. \\
\end{equation}
Here, $\partial \|X\|_1 = \{ H \in \mathbb{R}^{d \times m}~|~H_{ij}= {\rm sign} (X_{ij}) \text{ if } X_{ij} \neq 0, H_{ij} \in [-1,1] \text{ else}  \}$,
and $\Pi(a_{i}) := I_m - a_i a_i^{\top}$ is an orthogonal projection operator.

Note, that $\partial_2 f (X,A)$ is in fact a Riemannian gradient on the oblique manifold. Hence we may write $\partial_2 f (X,A)=\nabla_2 f (X,A)$. 
For the purpose of finding the subgradient with smallest Riemannian norm, each entry of $\partial_1 f (X,A)$ can be minimized independently. 
Let us denote the second summand at the right-hand side of \eqref{eq:17} by
\begin{equation}
B:=\sum\limits_{i=1}^{k} \lambda_{i}  {\mathcal D}^\top \Phi_{i}^\top \left(\Phi_{i} {\mathcal D} X a_{i} - \widehat{y}_i \right) a_i^\top 
\end{equation}
and define $C \in \mathbb{R}^{d \times m}$ as the matrix with $(i,j)$-entries
\begin{equation}
C_{ij}:= \left\{ \begin{array}{ll}
 {\rm sign}(X_{ij})  &\mbox{ if $X_{ij}\neq0$} \\[.2cm]
  -{\rm sign}(B_{ij}) \min\{ {|B_{ij}|, 1}\}  &\mbox{ otherwise.}
 \end{array} \right.
\end{equation}
Then the subgradient $G(X,A) \subset \partial f (X,A)$ having smallest Riemmanian norm is given explicitly by  
\begin{equation}
	G(A,X)=\left(C+B, \nabla_2 f (X,A) \right).
\end{equation}

\subsection{A Conjugate Subgradient Algorithm}
The development of a Riemannian Conjugate Subgradient method requires
the concept of parallel transport along 
a geodesic.
Since we consider the oblique manifold as a Riemannian submanifold of a product of unit spheres,
the formulas for the parallel transport as well as the exponential mapping read accordingly.

%
%
%
Let 
\begin{equation}
\label{eq:99}
	\tau_{x,\xi} (\psi) := \psi - \tfrac{\xi^{\top} \psi}{\| \xi \|^{2}} 
	\left( x \| \xi \| \sin t \| \xi \| + \xi (1 - \cos t \| 
	\xi \|) \right).
\end{equation}
be the parallel transport of a tangent vector $\psi \in 
T_{x}S^{m-1}$ along the great circle $\gamma_{x,\xi}(t)$ in the direction 
$\xi \in 
T_{x}S^{m-1}$ on the unit sphere $S^{m-1}$, where
\begin{equation}
\label{eq:15}
	\begin{split}
		\gamma_{x,\xi} \colon \mathbb{R} & \to S^{m-1}, \\
		\gamma_{x,\xi}(t) & := \left\{\!\! \begin{array}{ll}
			x, & \; \| \xi \| = 0; \\
			x \cos t \| \xi \|\! + \xi \tfrac{\sin t \| \xi\|}
			{\| \xi \|}, & \; \mathrm{otherwise}.
		\end{array} \right.
	\end{split}
\end{equation}
Then the parallel transport of $(H, \Psi) \in T_{(X,A)} \mathbb{R}^{d \times m} \times \mathcal{OB}(m, k)$ 
with respect to the Levi-Civita connection along the geodesic 
$\gamma_{(X,A)}$ in the 
direction $(Z,\Xi) \in T_{(X,A)} \mathbb{R}^{d \times m} \times 
\mathcal{OB}(m, k)$, i.e.
\begin{equation}
\label{eq:19}
	\begin{split}
		\gamma_{(X,A)} \colon \mathbb{R} & \to \mathbb{R}^{d \times m} 
		\times \mathcal{OB}(m, k), \\
		\gamma_{(X,A)}(t) & := \left( X+tZ, 
		[\gamma_{a_{i},\xi_{i}}(t)]_{i=1,\ldots,k}\right)
	\end{split}
\end{equation}
is given by
\begin{equation}
\label{eq:20}
	\tau_{(X,A),(Z,\Xi)}(H,\Psi) := \left(H, 
	\left[ \tau_{a_{i},\xi_{i}}(\psi_{i})\right]_{i=1,\ldots,k}\right).
\end{equation}

A conjugate subgradient algorithm 
for the problem as def{i}ned in 
\eqref{eq:optprob4} is summarized as follows. \vspace{-3mm}
\begin{center}
\begin{tabular}{p{0.97\columnwidth}}
	\toprule\hline\vspace*{-3mm}
	\hspace*{-0.5mm} \begin{algorithm}\label{algo:01}
	A conjugate subgradient CS-BSS algorithm
	\end{algorithm} \\[0.5mm]
	\hline\hline\vspace*{-1.5mm}
	\hspace*{-1mm} Step 1: Initialize $X^{(0)} 
		\in \mathbb{R}^{d \times m}$ and $A^{(0)} \in 
		\mathcal{OB}(m,k)$. \\[0.5mm]
	\hspace*{10mm} Set $i=0$. \\[0.5mm]
	\hspace*{-1mm} Step 2: Set $i=i+1$, let $X^{(i)} 
		= X^{(i-1)}$ and $A^{(i)} = A^{(i-1)}$, \\[0.5mm]
	\hspace*{10mm} and compute \\[1mm]
	\hspace*{25mm} $G^{(1)} = H^{(1)} = 
		-G(A^{(i)},X^{(i)})$. \\[1.5mm]
	\hspace*{-1mm} Step 3: For $j = 1, \ldots, d \cdot m + m(m-1) -1$: \\[1mm]
	\hspace*{6mm} ($i$) Update $\left(X^{(i)},A^{(i)}\right) \gets 
		\gamma_{\left(X^{(i)},A^{(i)}\right),H^{(i)}}
		\left(\lambda^{*}\right)$, \\[0.5mm] 
	\hspace*{10.5mm} where \\[1mm]
	\hspace*{25mm} $\lambda^{*} = \underset{\lambda\in\mathbb{R}}
		{\operatorname{argmin}}~\gamma_{\left(X^{(i)},A^{(i)}\right),H^{(i)}}
		\left(\lambda\right)$; \\[3.5mm]
	\hspace*{5mm} ($ii$) Compute $G^{(j+1)} = 
		-G(A^{(i)},X^{(i)})$; \\[1mm]
	\hspace*{4mm} ($iii$) Update $H^{(j+1)} \!\gets\!
		G^{(j+1)} + \mu~\tau_{(X^{(i)},A^{(i)}),
		H^{(j)}}(H^{(j)})$, \\[0.5mm]
	\hspace*{10.5mm} where $\mu$ is chosen according to \eqref{eq:dirupdate}; \\[1mm]
	\hspace*{5mm} ($iv$) If $\left\| H^{(j+1)}-H^{(j)} \right\|$ 
		is small enough, go to Step 2. \\[1mm]
	\hspace*{-1mm} Step 4: 
		If $\left\| A^{(i+1)}-A^{(i)} \right\| + \left\| 
		X^{(i+1)}-X^{(i)} \right\|$ is small \\[1mm] 
	\hspace*{10mm} enough, stop. Otherwise, go to Step 2. \\[1mm]
	\hline\bottomrule
\end{tabular}
\end{center}



\section{Numerical Results}
\label{sec:04}

\begin{figure*}[!ht]
\centering
\subfigure[Comparison in terms of Amari error]{
	\includegraphics[width=0.45\textwidth]{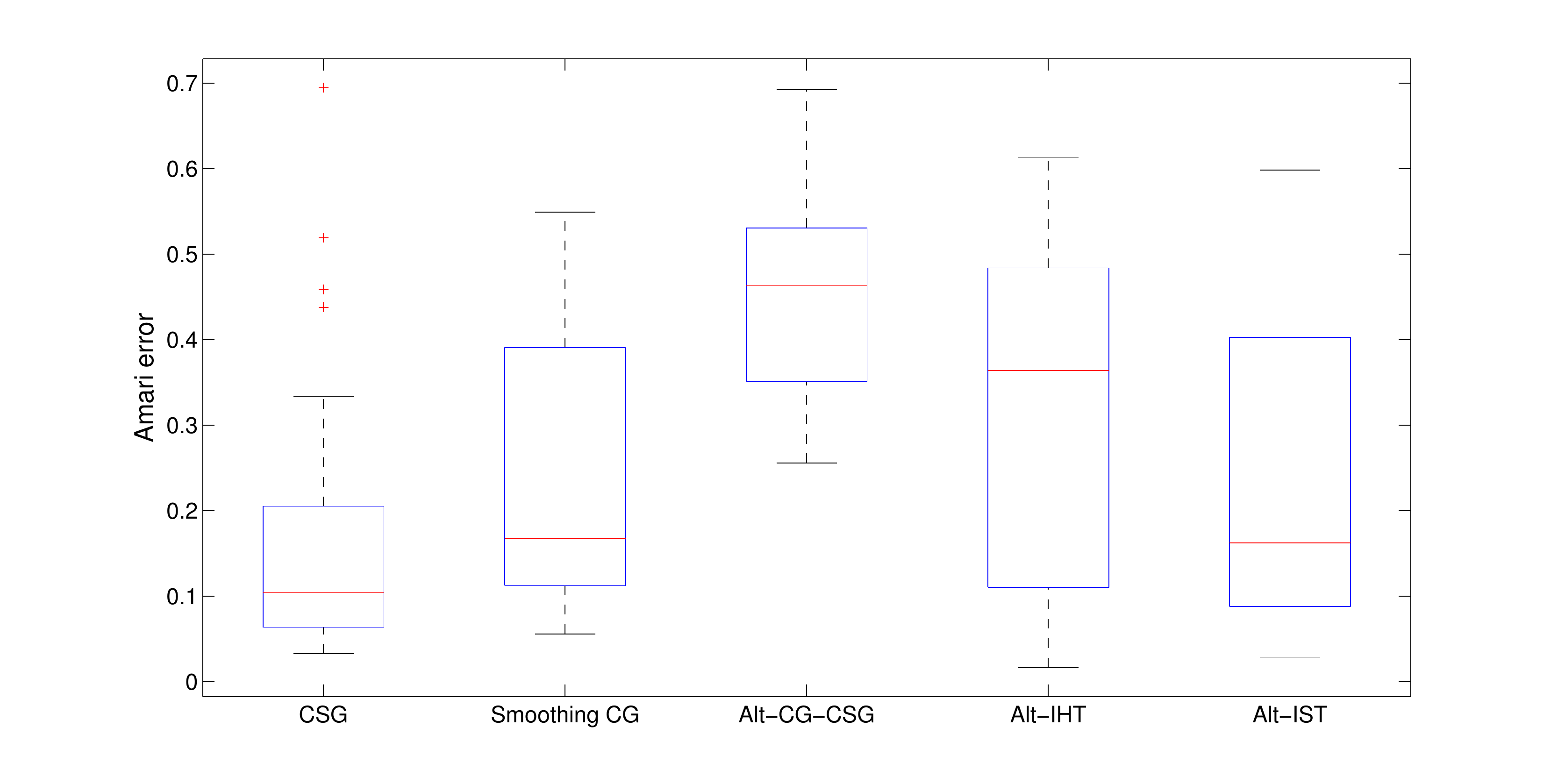}
	\label{fig:sub2}
}
\subfigure[Comparison in terms of SNR]{
	\includegraphics[width=0.45\textwidth]{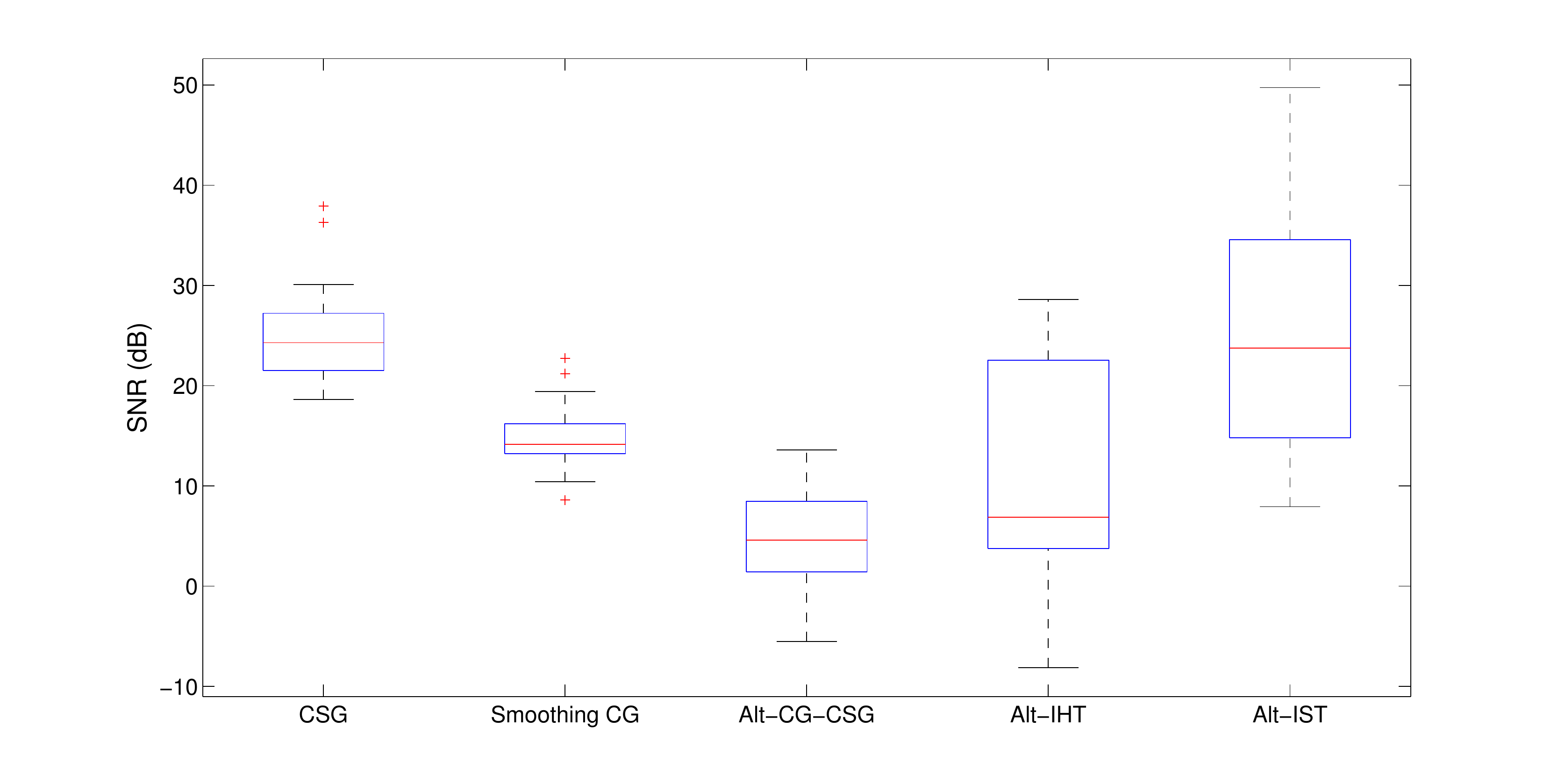}
	\label{fig:sub3}
}
\label{fig:01}
\caption{Experimental results.}\vspace*{-5mm}
\end{figure*}

%
In our experiment, we investigate the performance of the proposed CSG algorithm compared with several state of the art algorithms.
By noticing the similarity between our present work with the generalized 
MCA method in \cite{Bobin:07}, we adapt the alternating soft-thresholding 
based method and the hard thre\-sholding \cite{blumensath:09} (where $\ell_0$ 
is used as a sparsity measure) to our CS-BSS setting. We refer to them
as \emph{Alt-IST} and \emph{Alt-IHT}, respectively.
Similarly, we adapt the CSG algorithm in the alternating manner
(referred to as \emph{Alt-CG-CSG}).
Finally, one alternative common approach to deal with the nondifferentiability of the cost 
function is to employ a smooth approximation. 
We then refer to the standard CG algorithm with certain smooth approximation 
as \emph{smoothing CG}.


In the experiment, we generate three sources signals with $n = 1000$ samples.
The number of nonzero entries is equal to $100$, which are generated 
according to a Laplacian distribution. We randomly pick $300$ samples
as the compressively sensed mixtures.
The performance is measured in terms of both separation quality, measured by the Amari Error \cite{amar:nips96}, and
reconstruction quality in terms of Signal to Noise Ratio (SNR).
Figure~I illustrates the box plot of the results from 50 experiments of all methods.
Among all five methods, the 
\emph{Alt-CG-CSG} performs the worst on average in terms of both 
separation and reconstruction qualities, while the proposed CSG 
algorithm provides consistent and satisfactory results.
Finally, the small variances of all CG/CSG type algorithms in terms of reconstruction also suggest their reliable performance
compared with the thresholding based approaches.

\section{Conclusion}
\label{sec:05}
In this work, the authors propose a method for separating linearly mixed sparse signals from compressively sensed mixtures. The proposed approach allows each mixture to be sensed individually. If sampling takes place in the time domain, this means that the sensors do not have to be synchronized. The arising optimization problem is tackled with a conjugate subgradient type method, which is based on a new concept for optimizing non-differentiable functions under smooth constraints.



\begin{thebibliography}{10}

\bibitem{klei:spars11}
M.~Kleinsteuber and H.~Shen, ``Blind source separation of compressively sensed
  signals,'' in \emph{Proceedings of the $4^{th}$ Workshop on Signal Processing
  with Adaptive Sparse Structured Representations}, 2011, p.~80.

\bibitem{hayk:uaf1}
S.~Haykin, \emph{Unsupervised Adaptive Filtering}.\hskip 1em plus 0.5em minus
  0.4em\relax Wiley-Interscience, 2000, vol. 1: Blind Source Separation.

\bibitem{berm:ieeespl05}
S.~Bermejo, ``Finite sample effects in higher order statistics contrast
  functions for sequential blind source separation,'' \emph{IEEE Signal
  Processing Letters}, vol.~12, no.~6, pp. 481--484, 2005.

\bibitem{dono:ieeetit06}
D.~L. Donoho, ``Compressed sensing,'' \emph{IEEE Transactions on Information
  Theory}, vol.~52, pp. 1289--1306, 2006.

\bibitem{zibu:nc01}
M.~Zibulevsky and B.~A. Pearlmutter, ``Blind source separation by sparse
  decomposition in a signal dictionary,'' \emph{Neural Computation}, vol.~13,
  no.~4, pp. 863--882, 2001.

\bibitem{Li:2004}
Y.~Li, A.~Cichocki, and S.-i. Amari, ``Analysis of sparse representation and
  blind source separation,'' \emph{Neural Computation}, vol.~16, pp.
  1193--1234, 2004.

\bibitem{Elad:05}
M.~Elad, J.~Starck, P.~Querre, and D.~l. Donoho, ``Simultaneous cartoon and
  texture image inpainting using morphological component analysis ({MCA}),''
  \emph{Applied and Computational Harmonic Analysis}, vol.~19, no.~3, pp.
  340--358, 2005.

\bibitem{Bobin:07}
J.~Bobin, J.-L. Starck, J.~Fadili, and Y.~Moudden, ``Sparsity and morphological
  diversity in blind source separation,'' \emph{IEEE Transactions on Image
  Processing}, vol.~16, no.~11, pp. 2662 --2674, 2007.

\bibitem{cg:wolfe:1975}
P.~Wolfe, ``A method of conjugate subgradients for minimizing nondifferentiable
  functions,'' in \emph{Nondifferentiable Optimization}, ser. Mathematical
  Programming Studies, 1975, vol.~3, pp. 145--173.

\bibitem{hurley:09}
N.~Hurley and S.~Rickard, ``Comparing measures of sparsity,'' \emph{IEEE
  Transactions on Information Theory}, vol.~55, no.~10, pp. 4723 --4741, 2009.

\bibitem{bobi:tip07}
J.~Bobin, J.~Starck, J.~M. Fadili, Y.~Moudden, and D.~L. Donoho,
  ``Morphological component analysis: An adaptive thresholding strategy,''
  \emph{IEEE Transactions on Image Processing}, vol.~16, no.~11, pp.
  2675--2681, 2007.

\bibitem{wrig:ieeetsp09}
S.~J. Wright, R.~D. Nowak, and M.~A.~T. Figueiredo, ``Sparse reconstruction by
  separable approximation,'' \emph{IEEE Transactions on Signal Processing},
  vol.~57, pp. 2479--2493, 2009.

\bibitem{shen:lva10}
H.~Shen and M.~Kleinsteuber, ``Complex blind source separation via simultaneous
  strong uncorrelating transform,'' in \emph{Lecture Notes in Computer Science,
  Proceedings of the $9^{th}$ International Conference on Latent Variable
  Analysis and Signal Separation" (LVA/ICA 2010)}, vol. 6365.\hskip 1em plus
  0.5em minus 0.4em\relax Berlin/Heidelberg: Springer-Verlag, 2010, pp.
  287--294.

\bibitem{como:sp94}
P.~Comon, ``Independent component analysis, a new concept?'' \emph{Signal
  Processing}, vol.~36, no.~3, pp. 287--314, 1994.

\bibitem{ferreira:98}
O.~P. Ferreira and P.~R. Oliveira, ``Subgradient algorithm on {R}iemannian
  manifolds,'' \emph{Journal of Optimization Theory and Applications}, vol.~97,
  pp. 93--104, 1998.

\bibitem{ledyaev:07}
Y.~S. Ledyaev and Q.~J. Zhu, ``Nonsmooth analysis on smooth manifolds,''
  \emph{Transactions of the American Mathematical Society}, vol. 359, no.~8,
  pp. 3687--3732, 2007.

\bibitem{nocedal:2006}
J.~Nocedal and S.~J. Wright, \emph{{Numerical Optimization, 2nd Ed.}}\hskip 1em
  plus 0.5em minus 0.4em\relax New York: Springer, 2006.

\bibitem{absi:oamm08}
P.-A. Absil, R.~Mahony, and R.~Sepulchre, \emph{Optimization Algorithms on
  Matrix Manifolds}.\hskip 1em plus 0.5em minus 0.4em\relax Princeton, NJ:
  Princeton University Press, 2008.

\bibitem{blumensath:09}
T.~Blumensath and M.~E. Davies, ``Iterative hard thresholding for compressed
  sensing,'' \emph{Applied and Computational Harmonic Analysis}, vol.~27,
  no.~3, pp. 265--274, 2009.

\bibitem{amar:nips96}
S.~Amari, A.~Cichocki, and H.~H. Yang, ``A new learning algorithm for blind
  signal separation,'' \emph{Advances in Neural Information Processing
  Systems}, vol.~8, pp. 757--763, 1996.

\end{thebibliography}
\end{document}